\documentstyle[preprint,aps,prb,eqsecnum,tighten]{revtex}
\begin{document}
\preprint{cond-mat/9403034}
\draft
\title{Conductance fluctuations, weak localization, and shot noise\\
for a ballistic constriction in a disordered wire}
\author{C. W. J. Beenakker and J. A. Melsen}
\address{Instituut-Lorentz, University of Leiden\\
P.O. Box 9506, 2300 RA Leiden, The Netherlands}
\date{Submitted 8 March 1994}
\maketitle
\begin{abstract}
This is a study of phase-coherent conduction through a ballistic point contact
with disordered leads. The disorder imposes mesoscopic (sample-to-sample)
fluctuations and weak-localization corrections on the conductance, and also
leads to time-dependent fluctuations (shot noise) of the current. These effects
are computed by means of a mapping onto an unconstricted conductor with a
renormalized mean free path. The mapping holds both in the metallic and in the
localized regime, and permits a solution for arbitrary ratio of mean free path
to sample length. In the case of a single-channel quantum point contact, the
mapping is onto a one-dimensional disordered chain, for which the complete
distribution of the conductance is known. The theory is supported by numerical
simulations.
\end{abstract}
\newpage
\narrowtext
\section{Introduction}

The problem addressed in this paper is that of phase-coherent electron
transport through a ballistic point contact between disordered metals. The
geometry is shown schematically in Fig.\ \ref{constricted}. The same problem
was studied recently by Maslov, Barnes, and Kirczenow (MBK),\cite{Mas93} to
which paper we refer for an extensive introduction and bibliography. The
analytical theory of MBK is limited to the case that the mean free path $l$ for
elastic impurity scattering is much greater than the total length $L$ of the
system. In this ``quasi-ballistic'' case of $l\gg L$, the backscattering
through the point contact by disorder in the wide regions can be treated
perturbatively. In the present paper we go beyond MBK by solving the problem
for arbitrary ratio of $l$ and $L$, from the quasi-ballistic, through the
diffusive, into the localized regime of quantum transport.

Just as in Ref.\ \onlinecite{Mas93}, we model the scattering by the impurities
and by the constriction by {\em independent\/} and {\em isotropic\/} transfer
matrices. That is to say, we write the transfer matrix $M$ of the whole system
as the product $M=M_{2}M_{0}M_{1}$ of the transfer matrices $M_{1}$ and $M_{2}$
of the two wide disordered regions and the transfer matrix $M_{0}$ of the
ballistic constriction, and then we assume that the three transfer matrices are
distributed according to independent and isotropic distributions $p_{i}(M_{i})$
($i=0,1,2$). [A distribution $p(M)$ is called {\em isotropic\/} if it is only a
function of the eigenvalues of $MM^{\dagger}$.] The assumption of three
independent transfer matrices requires a spatial separation of scattering by
the impurities and by the constriction, which prevents us from treating the
effects of impurity scattering on the conductance quantization. (This problem
has been treated extensively in the past, cf.\ Ref.\ \onlinecite{Das93} for a
recent review.) The isotropy assumption for the transfer matrix $M_{0}$ of the
constriction is a simple but realistic model of the coupling between wide and
narrow regions, which implies that all $N$ transverse modes in the wide regions
(of width $W$) to the left and right of the constriction (of width $W_{0}$) are
equally coupled to each other.\cite{Sza89} The basic requirement here is that
the widening from $W_{0}$ to $W$ occurs abruptly and without spatial
symmetries.\cite{Note1} The isotropy assumption for the transfer matrices
$M_{1}$ and $M_{2}$ of the disordered regions (of length $L_{1}$ and $L_{2}$)
requires aspect ratios $L_{1}/W,L_{2}/W\gg 1$ corresponding to a wire
geometry.\cite{Sto91} Finally, we assume that the impurity scattering is weak
in the sense that $l\gg\lambda_{\rm F}$ (with $\lambda_{\rm F}$ the Fermi
wavelength). Under these assumptions we can treat the impurity scattering
within the framework of the Dorokhov-Mello-Pereyra-Kumar (DMPK)
equation.\cite{Dor82,Mel88}

The key result which enables us to go beyond MBK is a mapping between the
constricted and unconstricted geometries in Figs.\ \ref{constricted} and
\ref{unconstricted}. The unconstricted geometry of Fig.\ \ref{unconstricted} is
a disordered wire of length $L=L_{1}+L_{2}$, with $N_{0}$ transverse modes and
mean free path $l/\nu$. The number $N_{0}$ is determined by the quantized
conductance $N_{0}(2e^{2}/h)$ of the point contact in the constricted geometry.
The fraction $\nu$ is defined by
\begin{equation}
\nu=\frac{\beta N_{0}+2-\beta}{\beta N+2-\beta}.\label{nudef}
\end{equation}
(Here $\beta\in\{1,2,4\}$ is the Dyson index, which equals 1 in zero magnetic
field, 2 in a time-reversal-symmetry breaking magnetic field, and 4 in zero
field with strong spin-orbit scattering.) Starting from the DMPK equation, we
will deduce (in Sec.\ II) that the conductance has the {\em same\/} probability
distribution in the two geometries. The equivalence holds for {\em all\/}
moments of the conductance, so that it allows us to obtain (in Sec.\ III) the
effect of the point contact on weak localization and universal conductance
fluctuations directly from known results for disordered wires\cite{Mel91} ---
without the restriction $l\gg L$ of Ref.\ \onlinecite{Mas93}. It also holds for
other transport properties than the conductance (in fact it holds for the {\em
complete\/} distribution of the transmission eigenvalues). As an example of
current interest, we will compute the suppression of shot noise in the point
contact geometry.

In Sec.\ IV we consider the case $N_{0}=1$ of a quantum point contact with a
single transmitted channel. The mapping is then onto a single-mode wire (or
one-dimensional chain) of length $L$ and mean free path $\frac{1}{2}(\beta
N+2-\beta)l$. The 1D chain has been studied extensively in the past as the
simplest possible system exhibiting localization.\cite{Lan70,And80} The precise
correspondence with the problem of a single-channel ballistic constriction in a
multi-channel disordered wire seems to be both novel and unexpected. From this
correspondence we predict that the resistance $R$ of the point contact has an
exponential distribution,
\begin{equation}
{\cal P}(R)=\beta\frac{e^{2}}{h}\,\frac{Nl}{L}\exp\left[
-\beta\frac{e^{2}}{h}\,\frac{Nl}{L}(R-h/2e^{2})\right],\;\;R\geq
h/2e^{2},\label{PofR}
\end{equation}
provided the disordered wire is metallic ($Nl/L\gg 1$). The width of the
distribution decreases by a factor of two upon breaking time-reversal symmetry
in the absence of spin-orbit scattering ($\beta=1\rightarrow\beta=2$).

To test the theoretical predictions we present (in Sec.\ V) results of
numerical simulations, both for $N_{0}\gg 1$ and for $N_{0}=1$. The numerical
data for the density of transmission eigenvalues provides independent support
for the mapping. In particular, we find good agreement with Eq.\ (\ref{PofR}),
including the decrease in width upon application of a magnetic field.

\section{Mapping of constricted unto unconstricted geometry}

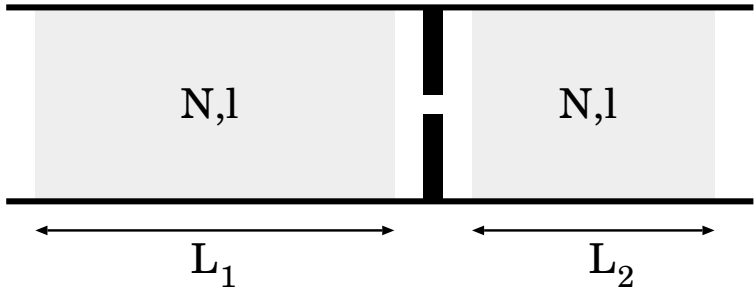
\begin{figure}
\caption[]{
Schematic illustration of the point contact geometry, consisting of a ballistic
constriction [with conductance $N_{0}(2e^2/h)$] in a disordered wire [with
length $L=L_{1}+L_{2}$, mean free path $l$, and $N$ transverse modes]. To
define a scattering geometry, the disordered regions (dotted) and the point
contact (black) are separated by scattering-free segments.
\label{constricted}}
\end{figure}

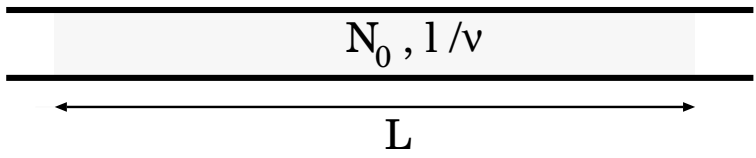
\begin{figure}
\caption[]{
Unconstricted geometry, with length $L$, mean free path $l/\nu$, and $N_{0}$
transverse modes. The key result of this paper is the equivalence with the
constricted geometry of Fig.\ \protect\ref{constricted}, for $\nu$ given by
Eq.\ (\protect\ref{nudef}).
\label{unconstricted}}
\end{figure}

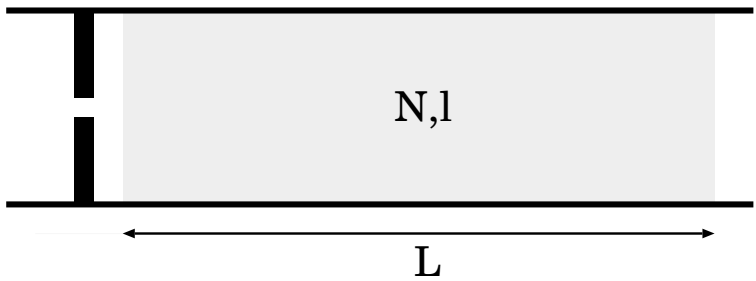
\begin{figure}
\caption[]{
Constricted geometry with all disorder at one side of the point contact. For
isotropic transfer matrices this geometry is statistically equivalent to that
of Fig.\ \protect\ref{constricted}.
\label{constricted2}}
\end{figure}

The first step is to show that the geometry of Fig.\ \ref{constricted}, with
lengths $L_{1}$ and $L_{2}$ of disordered wire to the left and right of the
point contact, is equivalent to the geometry of Fig.\ \ref{constricted2}, with
a length $L=L_{1}+L_{2}$ of disordered wire to one side only. The transfer
matrix for Fig.\ \ref{constricted} is $M=M_{2}M_{0}M_{1}$, the transfer matrix
of Fig.\ \ref{constricted2} is $M'=M_{0}M_{1}M_{2}$. The corresponding
probability distributions $p(M)$ and $p'(M')$ are
\begin{eqnarray}
p&=&p_{2}\ast p_{0}\ast p_{1},\label{pdef}\\
p'&=&p_{0}\ast p_{1}\ast p_{2},\label{pprimedef}
\end{eqnarray}
where the symbol $\ast$ denotes a convolution:
\begin{equation}
p_{i}\ast p_{j}(M)=\int\!dM_{j}\,p_{i}(MM_{j}^{-1})p_{j}(M_{j}).
\label{convolution}
\end{equation}
(The invariant measure $dM$ on the group of transfer matrices is introduced in
Refs.\ \onlinecite{Mel88} and \onlinecite{Mel91}.) Isotropic distributions have
the property that their convolution does not depend on the order: $p_{i}\ast
p_{j}=p_{j}\ast p_{i}$ if both $p_{i}$ and $p_{j}$ are isotropic (see the
Appendix for a proof). It follows that $p=p'$, and hence that the geometries of
Figs.\ \ref{constricted} and \ref{constricted2} are equivalent. Note that the
isotropy assumption is crucial here, otherwise the convolution would not
commute.

The second step is to show the equivalence of the constricted geometry of Fig.\
\ref{constricted2} with the unconstricted geometry of Fig.\
\ref{unconstricted}. We recall\cite{Sto91} that the $2N$ eigenvalues of the
transfer matrix product $MM^{\dagger}$ come in inverse pairs $\exp(\pm
2x_{n})$, $n=1,2,\ldots N$. The ratio $L/x_{n}\in[0,\infty)$ has the
significance of a channel-dependent localization length. We define
\begin{eqnarray}
T_{n}&=&1/\cosh^{2}x_{n},\label{Tdef}\\
\lambda_{n}&=&\sinh^{2}x_{n}=(1-T_{n})/T_{n}.\label{lambdadef}
\end{eqnarray}
The numbers $T_{n}\in[0,1]$ are the transmission eigenvalues (i.e.\ the
eigenvalues of the matrix product $tt^{\dagger}$, with $t$ the $N\times N$
transmission matrix). A ballistic point contact, with conductance
$N_{0}(2e^{2}/h)$, has to a good approximation $T_{n}=1$ ($\lambda_{n}=0$) for
$1\leq n\leq N_{0}$, and $T_{n}=0$ ($\lambda_{n}\rightarrow\infty$) for
$N_{0}+1\leq n\leq N$. (This is a statement about transmission {\em
eigenvalues}, not about the transmission probabilities of individual modes,
which are all of order $N_{0}/N$.) The joint probability distribution
$P_{N}(\lambda_{1},\lambda_{2},\ldots\lambda_{N},L)$ of the $\lambda$-variables
depends on the length $L$ of the disordered wire according to the DMPK
equation,\cite{Dor82,Mel88}
\begin{eqnarray}
&&\case{1}{2}l(\beta N+2-\beta)\frac{\partial P_{N}}{\partial L}=
\sum_{n=1}^{N}
\frac{\partial}{\partial\lambda_{n}}\lambda_{n}(1+\lambda_{n})
J_{N}\frac{\partial}{\partial\lambda_{n}}\frac{P_{N}}{J_{N}},\label{DMPK}\\
&&J_{N}=\prod_{i=1}^{N}\prod_{j=i+1}^{N}
|\lambda_{i}-\lambda_{j}|^{\beta}.\label{jacobian}
\end{eqnarray}
In this formulation the ballistic point contact appears as an initial condition
\begin{equation}
\lim_{L\rightarrow 0}P_{N}=\lim_{\Lambda\rightarrow\infty}
\prod_{n=1}^{N_{0}}\delta(\lambda_{n})\,\prod_{n=N_{0}+1}^{N}
\delta(\lambda_{n}-\Lambda).\label{Pinitial}
\end{equation}

The closed channels $N_{0}+1\leq n\leq N$ are irrelevant for conduction and can
be integrated out. The reduced distribution function
$\tilde{P}_{N}(\lambda_{1},\lambda_{2},\ldots\lambda_{N_{0}},L)$ is defined by
\begin{equation}
\tilde{P}_{N}=\int_{0}^{\infty}\!d\lambda_{N_{0}+1}
\int_{0}^{\infty}\!d\lambda_{N_{0}+2}\cdots
\int_{0}^{\infty}\!d\lambda_{N}\,P_{N},\label{Plambdadef}
\end{equation}
and satisfies the evolution equation
\begin{mathletters}
\label{DMPK2}
\begin{eqnarray}
&&\case{1}{2}l(\beta N+2-\beta)\frac{\partial\tilde{P}_{N}}{\partial L}=
\sum_{n=1}^{N_{0}}
\frac{\partial}{\partial\lambda_{n}}\lambda_{n}(1+\lambda_{n})
J_{N_{0}}\frac{\partial}{\partial\lambda_{n}}
\frac{\tilde{P}_{N}}{J_{N_{0}}},\label{DMPK2a}\\
&&\lim_{L\rightarrow 0}\tilde{P}_{N}=
\prod_{n=1}^{N_{0}}\delta(\lambda_{n}).\label{DMPK2b}
\end{eqnarray}
\end{mathletters}%
We now compare with the unconstricted geometry of Fig.\ \ref{unconstricted},
which consists of a wire with $N_{0}$ transverse modes, length $L$, and mean
free path $l/\nu$. The probability distribution
$P_{N_{0}}(\lambda_{1},\lambda_{2},\ldots\lambda_{N_{0}},L)$ for this geometry
is determined by
\begin{mathletters}
\label{DMPK3}
\begin{eqnarray}
&&\case{1}{2}(l/\nu)(\beta N_{0}+2-\beta)
\frac{\partial P_{N_{0}}}{\partial L}=\sum_{n=1}^{N_{0}}
\frac{\partial}{\partial\lambda_{n}}\lambda_{n}(1+\lambda_{n})
J_{N_{0}}\frac{\partial}{\partial\lambda_{n}}
\frac{P_{N_{0}}}{J_{N_{0}}},\label{DMPK3a}\\
&&\lim_{L\rightarrow 0}P_{N_{0}}=
\prod_{n=1}^{N_{0}}\delta(\lambda_{n}).\label{DMPK3b}
\end{eqnarray}
\end{mathletters}%
Comparison of Eqs.\ (\ref{DMPK2}) and (\ref{DMPK3}) shows that
$\tilde{P}_{N}=P_{N_{0}}$ if $\nu$ is given by Eq.\ (\ref{nudef}), as
advertized in the Introduction.

We will apply the mapping between constricted and unconstricted geometries to
study the distribution of transport properties $A$ of the form
$A=\sum_{n}a(\lambda_{n})$, with $\lim_{\lambda\rightarrow\infty}a(\lambda)=0$
(so that only the channels $n\leq N_{0}$ contribute). We denote by ${\cal
P}(A,s)$ and ${\cal P}_{0}(A,s)$ the distribution of $A$ in, respectively, the
constricted and unconstricted geometries, with $s=L/$mean~free~path. Since the
mean free path in the constricted geometry is a factor $\nu$ smaller than in
the unconstricted geometry, we conclude that
\begin{equation}
{\cal P}(A,s)={\cal P}_{0}(A,\nu s).\label{key}
\end{equation}
This is the key result which allows us to solve the problem of a ballistic
constriction in a disordered wire, for arbitrary ratio $s$ of wire length to
mean free path.

\section{Many-channel point contact}

In this Section we study a point contact which has a conductance much greater
than $e^{2}/h$, so that $N_{0}\gg 1$. We mainly consider the metallic regime
$Nl/L\gg 1$, in which the conductance of the disordered region separately is
also much greater than $e^{2}/h$. Two transport properties are studied in
detail: Firstly the conductance $G$, given by the Landauer formula
\begin{equation}
G=G_{0}\sum_{n}T_{n},\label{Landauer}
\end{equation}
where $G_{0}=2e^{2}/h$ is the conductance quantum. Secondly the shot-noise
power $S$, given by\cite{But90}
\begin{equation}
S=S_{0}\sum_{n}T_{n}(1-T_{n}),\label{Buettiker}
\end{equation}
with $S_{0}=2e|V|G_{0}$ for an applied voltage $V$. We also study the
transmission-eigenvalue density, from which other transport properties can be
computed. In each case we apply the mapping (\ref{key}) between the constricted
and unconstricted geometries. The fraction $\nu$ which rescales the mean free
path in this mapping has, according to Eq.\ (\ref{nudef}), the series expansion
\begin{equation}
\nu=\frac{1}{N}\left[N_{0}-(1-2/\beta)(1-N_{0}/N)+
{\cal O}(N_{0}^{-1})\right].\label{nuseries}
\end{equation}
To lowest order, $\nu=N_{0}/N$. The next term, proportional to $1-2/\beta$,
contributes to the weak localization effect.

\subsection{Weak localization and conductance fluctuations}

The mean $\bar{G}$ and variance ${\rm Var}\,G$ of the conductance distribution
${\cal P}_{0}(G,\nu s)$ in the unconstricted geometry were computed by Mello
and Stone [Eq.\ (C23) in Ref.\ \onlinecite{Mel91}]
\begin{eqnarray}
&&\bar{G}/G_{0}=\frac{N_{0}}{1+\nu s}+\frac{1}{3}(1-2/\beta)\left(\frac{\nu
s}{1+\nu s}\right)^{3}+{\cal O}(\nu s/N_{0}),\label{Gbar0}\\
&&{\rm Var}\,G/G_{0}=\frac{2}{15}\beta^{-1}\left(1-\frac{1+6\nu s}{(1+\nu
s)^{6}}\right)+{\cal O}(\nu s/N_{0}).\label{VarG0}
\end{eqnarray}
Substitution of the expansion (\ref{nuseries}) yields for the constricted
geometry the average conductance $\bar{G}=G_{\rm series}+\delta G$, with
$G_{\rm series}$ given by $G_{\rm series}=G_{0}(N_{0}^{-1}+s/N)^{-1}$ and
$\delta G$ given by (denoting $\gamma\equiv N_{0}s/N$):
\begin{equation}
\delta G/G_{0}=(1-2/\beta)\left[
\frac{1}{3}\left(\frac{\gamma}{1+\gamma}\right)^{3}+
\left(1-\frac{N_{0}}{N}\right)\frac{\gamma}{(1+\gamma)^{2}}\right]+{\cal
O}(s/N).\label{deltaG}
\end{equation}
The term of order $s/N=L/Nl$ can be neglected in the metallic regime. The term
$G_{\rm series}$ is the series addition of the Sharvin conductance $G_{\rm
Sharvin}=G_{0}N_{0}$ of the ballistic point contact and the Drude
conductance\cite{Note2} $G_{\rm Drude}=G_{0}Nl/L$ of the disordered region. The
term $\delta G$ is the weak localization correction to the classical series
conductance. This term depends on the ratio $\gamma$ of the Sharvin and Drude
conductances as well as on the ratio $N_{0}/N$ of the width of the point
contact and the wide regions. In the limit $N_{0}/N\rightarrow 0$ at constant
$\gamma$, Eq.\ (\ref{deltaG}) simplifies to
\begin{equation}
\delta G/G_{0}=\frac{1}{3}(1-2/\beta)[1-(1+\gamma)^{-3}].\label{deltaG2}
\end{equation}
The variance ${\rm Var}\,G$ of the sample-to-sample fluctuations of the
conductance around the average depends only on $\gamma$ (to order $s/N$). From
Eqs.\ (\ref{nuseries}) and (\ref{VarG0}) we find
\begin{equation}
{\rm Var}\,G/G_{0}=\frac{2}{15}\beta^{-1}\left(1-
\frac{1+6\gamma}{(1+\gamma)^{6}}\right).\label{VarG}
\end{equation}

In Fig.\ \ref{vargplot} we have plotted $\delta G$ and $({\rm Var}\,G)^{1/2}$
as a function of $\gamma=G_{\rm Sharvin}/G_{\rm Drude}$. (The limit
$N_{0}/N\rightarrow 0$ is assumed for $\delta G$.) For large $\gamma$ the
curves tend to $\delta G_{\infty}=\frac{1}{3}(1-2/\beta)G_{0}$ and ${\rm
Var}\,G_{\infty}=\frac{2}{15}\beta^{-1}G_{0}^{2}$, which are the familiar
values\cite{Mel91} for weak localization and universal conductance fluctuations
in a wire geometry without a point contact. These values are universal to the
extent that they are independent of wire length and mean free path. The
presence of a point contact breaks this universality, but only if the Sharvin
conductance is smaller than the Drude conductance. For $\gamma>1$ the
universality is quickly restored, according to
\begin{eqnarray}
\frac{\delta G}{\delta G_{\infty}}&=&1-\gamma^{-3}+{\cal
O}(\gamma^{-4}),\label{deltaGasymp1}\\
\frac{{\rm Var}\,G}{{\rm Var}\,G_{\infty}}&=&1-6\gamma^{-5}+{\cal
O}(\gamma^{-6}).\label{VarGasymp1}
\end{eqnarray}
For $\gamma<1$ both $\delta G$ and ${\rm Var}\,G$ are suppressed by the
presence of the point contact, according to
\begin{eqnarray}
\delta G/G_{0}&=&(1-2/\beta)\gamma+{\cal O}(\gamma^{2}), \label{deltaGasymp2}\\
{\rm Var}\,G/G_{0}&=&2\beta^{-1}\gamma^{2}+{\cal
O}(\gamma^{3}).\label{VarGasymp2}
\end{eqnarray}

Maslov, Barnes, and Kirczenow\cite{Mas93} have studied the quasi-ballistic
regime $l\gg L$. They consider a geometry as in Fig.\ \ref{constricted}, with
$L_{1}=L_{2}$, and relate the variance ${\rm Var}\,G$ of the whole system to
the variance ${\rm Var}\,G_{1}$ of one of the two disordered segments of length
$\case{1}{2}L$. Their result (in the present notation) is
\begin{equation}
{\rm Var}\,G=\gamma^{2}(l/L_{1})^{2}\,{\rm Var}\,G_{1},\label{rmsG}
\end{equation}
in precise agreement with our small-$\gamma$ result (\ref{VarGasymp2}) [since
${\rm Var}\,G_{1}=2\beta^{-1}(L_{1}/l)^{2}$ for $l\gg L_{1}$].

So far we have considered the metallic regime $N/s\gg 1$. We now briefly
discuss the insulating regime $N/s\ll 1$. In the unconstricted geometry the
conductance then has a log-normal distribution,\cite{Sto91,Pic91}
\begin{equation}
{\cal P}_{0}(G,\nu s)=C\exp\left(-\frac{(2\beta^{-1}\nu s/N_{0}+\ln
G/G_{0})^{2}}{8\beta^{-1}\nu s/N_{0}}\right),\;\;{\rm if}\;\;\nu s/N_{0}\gg
1,\label{lnG0}
\end{equation}
with $C$ a normalization constant. The mapping (\ref{key}) implies that the
conductance in the constricted geometry has also a log-normal distribution,
with mean $\langle\ln G/G_{0}\rangle=-2\beta^{-1}s/N$ and variance ${\rm
Var}(\ln G/G_{0})=4\beta^{-1}s/N$. This distribution is independent of the
conductance of the point contact, as long as $N_{0}\gg 1$.

\subsection{Suppression of shot noise}

The average shot-noise power in the unconstricted geometry is [Eq.\ (A10) in
Ref.\ \onlinecite{Jon92}]
\begin{equation}
\bar{S}/S_{0}=\frac{1}{3}N_{0}(1+\nu s)^{-1}[1-(1+\nu s)^{-3}]+{\cal
O}(1).\label{Sbar0}
\end{equation}
The term ${\cal O}(1)$ is the weak localization correction on the shot noise,
which is not considered here. The mapping (\ref{key}) implies for the
constricted geometry
\begin{equation}
\bar{S}/S_{0}=\frac{1}{3}N_{0}(1+\gamma)^{-1} [1-(1+\gamma)^{-3}],\label{Sbar}
\end{equation}
with $\gamma\equiv N_{0}s/N$. Since $S_{0}N_{0}(1+\gamma)^{-1}=2e|V|G_{\rm
series}=2e|I|$ (with $I$ the current through the point contact), we can write
Eq.\ (\ref{Sbar}) in terms of the Poisson noise $S_{\rm Poisson}=2e|I|$,
\begin{equation}
\bar{S}=\frac{1}{3}S_{\rm Poisson}[1-(1+\gamma)^{-3}].\label{Sbar2}
\end{equation}

The suppression of the shot-noise power below the value $S_{\rm Poisson}$ of a
Poisson process is plotted in Fig.\ \ref{splot}, as a function of the ratio
$\gamma$ of Sharvin and Drude conductances. For $\gamma\ll 1$ the shot noise is
zero, as expected for a ballistic constriction.\cite{But90,Kul84,Khl87,Les89}
For $\gamma\gg 1$ the shot noise is one third the Poisson noise, as expected
for a diffusive conductor.\cite{Jon92,Bee92,Nag92} The formula (\ref{Sbar2})
describes the crossover between these two regimes.

\subsection{Density of transmission eigenvalues}

We consider the eigenvalue densities
\begin{eqnarray}
\rho(x,s)&=&\left\langle\sum_{n=1}^{N_{0}}
\delta(x-x_{n})\right\rangle,\label{rhoxdef}\\
\rho(T,s)&=&\left\langle\sum_{n=1}^{N_{0}}
\delta(T-T_{n})\right\rangle,\label{rhoTdef}
\end{eqnarray}
which are related by $\rho(T,s)=\rho(x,s)|dT/dx|^{-1}$ (with $T=1/\cosh^{2}x$).
The (irrelevant) closed channels $n>N_{0}$ have been excluded from the
densities. In the unconstricted geometry we have, according to Ref.\
\onlinecite{Bee94},
\begin{equation}
\rho(x,\nu s)=\frac{2}{\pi}N_{0}\,{\rm Im}\,U(x-{\rm i}0^{+},\nu s)+{\cal
O}(1),\label{rhox0}
\end{equation}
where the complex function $U(z,s)$ is determined by
\begin{equation}
U={\rm cotanh}\,(z-sU),\;\;0>{\rm Im}\,(z-sU)>-\case{1}{2}\pi.\label{Udef}
\end{equation}
The mapping (\ref{key}) implies for the constricted geometry
\begin{equation}
\rho(x,s)=\frac{2}{\pi}N_{0}\,{\rm Im}\,U(x-{\rm i}0^{+},N_{0}s/N)+{\cal
O}(1).\label{rhoxU}
\end{equation}

The solution $\rho(x,s)$ of Eqs.\ (\ref{Udef}) and (\ref{rhoxU}) is plotted in
Fig.\ \ref{rhoxplot}, for several values of $\gamma\equiv N_{0}s/N$. The inset
shows the corresponding density of transmission eigenvalues $\rho(T,s)$. For
$\gamma\lesssim 1$, $\rho(T,s)$ has a single peak at unit transmission. For
$\gamma\gtrsim 1$ a second peak develops near zero transmission, so that the
distribution becomes {\em bimodal}. A crossover from unimodal to bimodal
distribution on increasing the disorder has also been found in the case of a
tunnel barrier.\cite{Bee94,Naz94} The difference with a point contact is that
for a tunnel barrier the single peak is near zero, rather than near unit,
transmission.

\section{Single-channel point contact}

In this Section we study a point contact with a quantized conductance of
$2e^{2}/h$, so that $N_{0}=1$. The DMPK equation (\ref{DMPK3}) for the
distribution $P_{1}(\lambda_{1},L)\equiv P(\lambda,L)$ of the single
transmitted channel is
\begin{mathletters}
\label{DMPK4}
\begin{eqnarray}
&&(l/\nu)
\frac{\partial}{\partial L}P(\lambda,L)=
\frac{\partial}{\partial\lambda}\lambda(1+\lambda)
\frac{\partial}{\partial\lambda}P(\lambda,L),\label{DMPK4a}\\
&&\lim_{L\rightarrow 0}P(\lambda,L)=\delta(\lambda),\label{DMPK4b}
\end{eqnarray}
\end{mathletters}%
since $J_{1}\equiv 1$. Eq.\ (\ref{nudef}) for the fraction $\nu$ which rescales
the mean free path becomes
\begin{equation}
\nu=\frac{2}{\beta N+2-\beta}.\label{nu1def}
\end{equation}
The partial differential equation (\ref{DMPK4}) has been studied as early as
1959 in the context of propagation of radio-waves through a waveguide with a
random refractive index.\cite{Ger59,Pap71} In the eighties it was rederived and
investigated in great detail,\cite{Mel81,Abr81,Kir84,Kum85,Mel86} in connection
with the problem of localization in a 1D chain.\cite{Lan70,And80} The solution
can be written in terms of Legendre functions, or more conveniently in the
integral representation
\begin{equation}
P(\lambda,L)=(2\pi)^{-1/2}(\nu L/l)^{-3/2}{\rm e}^{-\nu L/4l}\int_{{\rm
arccosh}(1+2\lambda)}^{\infty}du\frac{u\exp(-u^{2}l/4\nu L)}{(\cosh
u-1-2\lambda)^{1/2}}.\label{Plambdasol}
\end{equation}

According to the Landauer formula (\ref{Landauer}), the conductance $G$ of the
whole system is related to the variable $\lambda\equiv(1-T)/T$ by
$G=G_{0}(1+\lambda)^{-1}$ (with $G_{0}=2e^{2}/h$). It follows that the
resistance $\delta R=1/G-h/2e^{2}$ after subtraction of the contact resistance
is just given by $\delta R=\lambda/G_{0}$. In view of the mapping (\ref{key}),
the resistance distribution ${\cal P}(\delta R,s)$ is given by
\begin{equation}
{\cal P}(\delta R,s)=G_{0}(2\pi)^{-1/2}(\nu s)^{-3/2}{\rm e}^{-\nu
s/4}\int_{{\rm arccosh}(1+2G_{0}\delta R)}^{\infty}du\frac{u\exp(-u^{2}/4\nu
s)}{(\cosh u-1-2G_{0}\delta R)^{1/2}}.\label{PdeltaR}
\end{equation}
The mean and variance of $\delta R$ can be computed either by integrating the
distribution (\ref{PdeltaR}), or directly from the differential equation
(\ref{DMPK4}).\cite{Mel81} The result is
\begin{eqnarray}
\overline{\delta R}&=&\frac{1}{2G_{0}}\left({\rm e}^{2\nu
s}-1\right),\label{deltaRbar}\\
{\rm Var}\,{\delta R}&=&\frac{1}{6G_{0}^{2}}\left({\rm e}^{6\nu
s}-\case{3}{2}{\rm e}^{4\nu s}+\case{1}{2}\right).\label{VardeltaR}
\end{eqnarray}

These results hold in both the metallic and the insulating regimes. We now
consider in some more detail the metallic regime $N/s\gg 1$. This implies $\nu
s\ll 1$. Eqs.\ (\ref{deltaRbar}) and (\ref{VardeltaR}) reduce to
\begin{equation}
\overline{\delta R}=\frac{2s}{G_{0}}(\beta N+2-\beta)^{-1}+{\cal
O}(s/N)^{2}=({\rm Var}\,{\delta R})^{1/2}.\label{barVardeltaR}
\end{equation}
The complete distribution of the resistance $\delta R$ (which follows from Eq.\
(\ref{PdeltaR}) in the limit $\nu s\ll 1$) is the exponential distribution
\begin{equation}
{\cal P}(\delta R,s)=\frac{G_{0}}{\nu s}\exp\left(-\frac{G_{0}}{\nu s}\delta
R\right),\;\delta R\geq 0.\label{Pexponential}
\end{equation}
For $N\gg 1$ the width $\nu s\simeq 2s/\beta N$ of the distribution
(\ref{Pexponential}) has the $1/\beta$ dependence announced in the Introduction
[Eq.\ (\ref{PofR})]. In Fig.\ \ref{abrikosov} we have plotted the exact
distribution (\ref{PdeltaR}) [solid curves] for several values of $s$ and
compared with the metallic limit (\ref{Pexponential}) [dashed curves]. For $\nu
s\lesssim 0.1$ [curves labeled a] the two results are almost indistinguishable.

To make connection with some of the recent literature, we remark that the
exponential resistance distribution (\ref{Pexponential}) implies for the
conductance the distribution
\begin{equation}
{\cal P}(G,s)=\frac{G_{0}}{\nu s}\,G^{-2}
\exp\left(\frac{1-G_{0}/G}{\nu s}\right),\;0\leq G\leq
G_{0},\label{Pexponential2}
\end{equation}
which is strongly peaked at $G=G_{0}$. This is completely different from the
conductance distribution of a quantum dot which is weakly coupled by two point
contacts to electron reservoirs.\cite{Pri93,Jal94}

\section{Numerical simulations}

To test the analytical predictions we have carried out numerical simulations of
the Anderson model in the geometry of Fig.\ \ref{constricted2}, using the
recursive Green's function technique.\cite{Bar91} The disordered region
(dotted) was modeled by a tight-binding Hamiltonian on a square lattice
(lattice constant $a$), with a random impurity potential at each site
(uniformly distributed between $\pm\case{1}{2}U_{\rm d}$). The constriction was
introduced by assigning a large potential energy to sites at one end of the
lattice (black in Fig.\ \ref{constricted}), so as to create a nearly
impenetrable barrier with an opening in the center. The constriction itself
contained no disorder (the disordered region started at two sites from the
barrier). The Fermi energy was chosen at $E_{\rm F}=1.5\,u_{0}$ from the band
bottom (with $u_{0}\equiv\hbar^{2}/2ma^{2}$). The ratio $s$ of sample length to
mean free path which appears in the theory was computed numerically from ${\rm
Tr}\,t_{\rm d}^{\vphantom{\dagger}}t_{\rm d}^{\dagger}=N(1+s)^{-1}$, with
$t_{\rm d}$ the transmission matrix of the disordered region without the
constriction.\cite{Note3}

The simulations for the many-channel and single-channel point contact are
discussed in two separate sub-sections.

\subsection{Many-channel point contact}

Two geometries were considered for the wide disordered region: a square
geometry ($L=W=285\,a$, corresponding to $N=119$), and a rectangular geometry
($L=285\,a$, $W=93\,a$, corresponding to $N=39$). In each case the width of the
constriction was $\frac{1}{3}W$ (corresponding to $N_{0}=40$ and $N_{0}=13$ in
the square and rectangular geometries, respectively). The length of the
constriction was one site. The strength $U_{\rm d}$ of the impurity potential
was varied between $0$ and $1.5\,u_{0}$, corresponding to $s$ between $0$ and
$11.7$.

In Fig.\ \ref{rhoxnumplot} we compare the integrated eigenvalue density
$N_{0}^{-1}\int_{0}^{x}dx'\,\rho(x',s)$, which is the quantity following
directly from the simulation. The points are raw data from a single sample.
(Sample-to-sample fluctuations are small, because the $x_{n}$'s are
self-averaging quantities.\cite{Sto91}) The data is in good agreement with the
analytical result of Sec.\ IIIC, {\em without any adjustable parameters}. No
significant geometry dependence was found (compare open and closed symbols in
Fig.\ \ref{rhoxnumplot}).

\subsection{Single-channel point contact}

We considered a square geometry ($L=W=47\,a$, corresponding to $N=20$), and a
rectangular geometry ($L=47\,a$, $W=23\,a$, corresponding to $N=10$). The point
contact was three sites wide and two sites long, corresponding to $N_{0}=1$.
(The conductance in the absence of disorder was within 5\% of $2e^{2}/h$.) The
distribution $P(\delta R,s)$ of the resistance $\delta R\equiv R-h/2e^{2}$ was
computed by collecting data for some $10^{4}$ realizations of the impurity
potential. To compare the cases $\beta=1$ and $\beta=2$, we repeated the
simulations in the presence of a magnetic field of 50 flux quanta $h/e$ through
the disordered region. (The magnetic field was graded to zero in the ideal
leads.) Two disorder strengths were considered: $U_{\rm d}=1.5\,u_{0}$
(corresponding to $s=1.8$) and $U_{\rm d}=3.0\,u_{0}$ (corresponding to
$s=8.3$). The results are collected in Fig.\ \ref{pnumplot} and are in good
agreement with the theoretical prediction (\ref{PdeltaR}), again without any
adjustable parameters. The theory agrees comparably well with the simulations
for the square and rectangular geometries, which shows that the condition $L\gg
W$ for the validity of the DMPK equation can be relaxed to a considerable
extent.

We find it altogether quite remarkable that the amusingly simple mapping
(\ref{key}) between the constricted and unconstricted geometries is capable of
reliably predicting the {\em complete\/} distribution of the point-contact
resistance, including the effect of broken time-reversal symmetry. We know of
no more conventional theoretical technique which could do the same.

\acknowledgments
Valuable discussions with M. J. M. de Jong, D. L. Maslov, and B. Rejaei are
gratefully acknowledged. This research was supported by the ``Ne\-der\-land\-se
or\-ga\-ni\-sa\-tie voor We\-ten\-schap\-pe\-lijk On\-der\-zoek'' (NWO) and by
the ``Stich\-ting voor Fun\-da\-men\-teel On\-der\-zoek der Ma\-te\-rie''
(FOM).

\appendix
\section*{Isotropically distributed transfer matrices commute}

We wish to show that the probability distribution $p(M)$ of a product
$M=M_{1}M_{2}M_{3}\cdots\,$ of transfer matrices is independent of the order of
the matrix product, under the assumption that the matrices $M_{i}$ are
independently distributed with isotropic distributions $p_{i}$. Since
$p=p_{1}\ast p_{2}\ast p_{3}\cdots$, it is sufficient to show that the
convolution
\begin{equation}
p_{i}\ast p_{j}(M)=\int\!dM_{j}\,p_{i}(MM_{j}^{-1})p_{j}(M_{j}) \label{pijdef}
\end{equation}
of any two isotropic distributions commutes.

By definition, the distribution $p(M)$ is isotropic if it is only a function of
the eigenvalues of the product $MM^{\dagger}$. This implies that $p(M)=p(M^{\rm
T})$. (The superscripts $\dagger$ and T denote, respectively, the hermitian
conjugate and the transpose of a matrix.) As shown by Mello et
al.,\cite{Mel88,Mel91} the convolution of two isotropic distributions is again
isotropic. Hence
\begin{eqnarray}
p_{i}\ast p_{j}(M)&=&p_{i}\ast p_{j}(M^{\rm T})\nonumber\\
&=&\int\!dM_{j}\,p_{i}(M^{\rm T}M_{j}^{-1})p_{j}(M_{j})\nonumber\\
&=&\int\!dM_{j}\,p_{i}({M_{j}^{\rm T}}^{-1}M)p_{j}(M_{j}^{\rm T})\nonumber\\
&=&\int\!dM_{i}\,p_{i}(M_{i})p_{j}(MM_{i}^{-1})\nonumber\\
&=&p_{j}\ast p_{i}(M),\label{commute}
\end{eqnarray}
which proves the commutativity of the convolution of isotropic distributions.

\begin{figure}
\caption[]{
Suppression by the point contact of the weak localization correction $\delta G$
and the root-mean-square conductance fluctuations $({\rm Var}\,G)^{1/2}$. The
dashed and solid curves are from Eqs.\ (\protect\ref{deltaG2}) and
(\protect\ref{VarG}), respectively. For $\gamma=G_{\rm Sharvin}/G_{\rm
Drude}=N_{0}s/N\gg 1$ the curves approach the values $\delta G_{\infty}$ and
$({\rm Var}\,G_{\infty})^{1/2}$ of an unconstricted disordered wire (normalized
to unity in the plot).
\label{vargplot}}
\end{figure}

\begin{figure}
\caption[]{
Suppression of the shot-noise power $S$ below the Poisson noise $S_{\rm
Poisson}$. The solid curve is computed from Eq.\ (\protect\ref{Sbar2}). The
one-third suppression of a diffusive conductor is indicated by the dashed line.
\label{splot}}
\end{figure}

\begin{figure}
\caption[]{
Density $\rho(x,s)$ as a function of $x$, computed from Eqs.\
(\protect\ref{Udef}) and (\protect\ref{rhoxU}) for several values of
$\gamma\equiv N_{0}s/N$. Curves a, b, c, d, and e correspond, respectively, to
$\gamma=0.2$, 0.5, 1, 2, and 4. The inset shows the corresponding density
$\rho(T,s)=\rho(x,s)|dT/dx|^{-1}$ of transmission eigenvalues $T=1/\cosh^{2}x$.
Note the crossover from unimodal to bimodal distribution near $\gamma=1$.
\label{rhoxplot}}
\end{figure}

\begin{figure}
\caption[]{
Probability distribution of the resistance $\delta R=R-h/2e^{2}$ of a
single-channel point contact, for several values of $\nu s=2(L/l)(\beta
N+2-\beta)^{-1}$. Curves a, b, and c, correspond, respectively, to $\nu s=0.1$,
0.2, and 0.5. The solid curves are computed from Eq.\ (\protect\ref{PdeltaR}),
the dashed curves are the exponential distribution (\protect\ref{Pexponential})
which is approached in the metallic regime $\nu s\ll 1$.
\label{abrikosov}}
\end{figure}

\begin{figure}
\caption[]{
Comparison between theory and simulation of the integrated eigenvalue density
for $N_{0}/N=1/3$ and for three different disorder strengths ($s=0,3,11.7$).
Solid curves are from Eqs.\ (\protect\ref{Udef}) and (\protect\ref{rhoxU}),
data points are the  $N_{0}$ smallest $x_{n}$'s from the simulation plotted in
ascending order versus $n/N_{0}$ [filled data points are for a square geometry,
open points for a rectangular disordered region ($L/W=3$)].
\label{rhoxnumplot}}
\end{figure}

\begin{figure}
\caption[]{
Comparison between theory and simulation of the distribution of the excess
resistance $\delta R$ of a single-channel point contact, for $s=1.8$ (a) and
$s=8.3$ (b). The histograms are the numerical data (for square and rectangular
disordered regions), the smooth curves are computed from Eq.\
(\protect\ref{PdeltaR}) --- without any adjustable parameters. Solid curves are
for zero magnetic field ($\beta=1$), dash-dotted curves for a magnetic flux of
$50\,h/e$ through the disordered region ($\beta=2$). For clarity, the curves
for the square geometry are offset vertically by 1.5 and 0.25 in figures (a)
and (b), respectively.
\label{pnumplot}}
\end{figure}

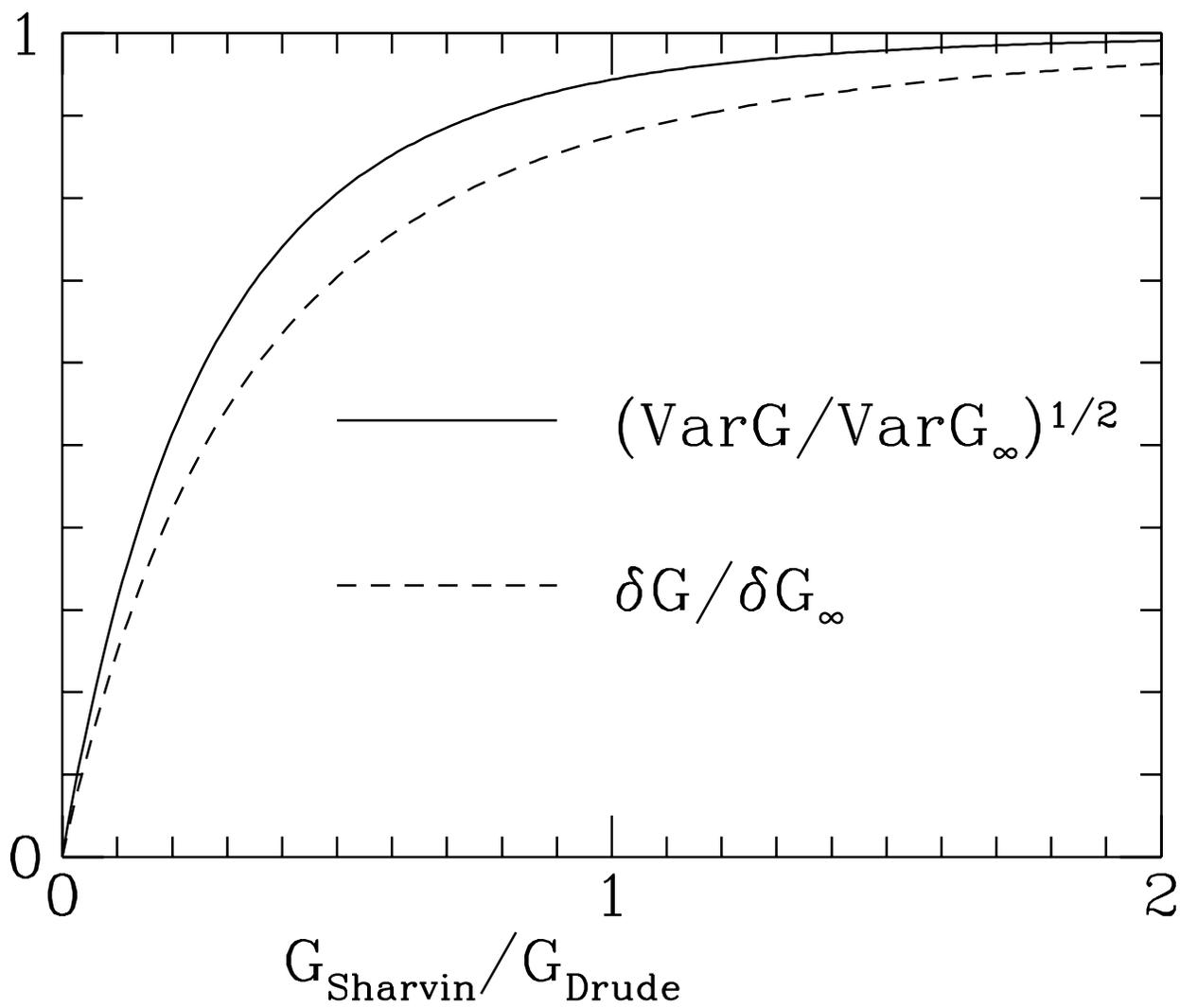
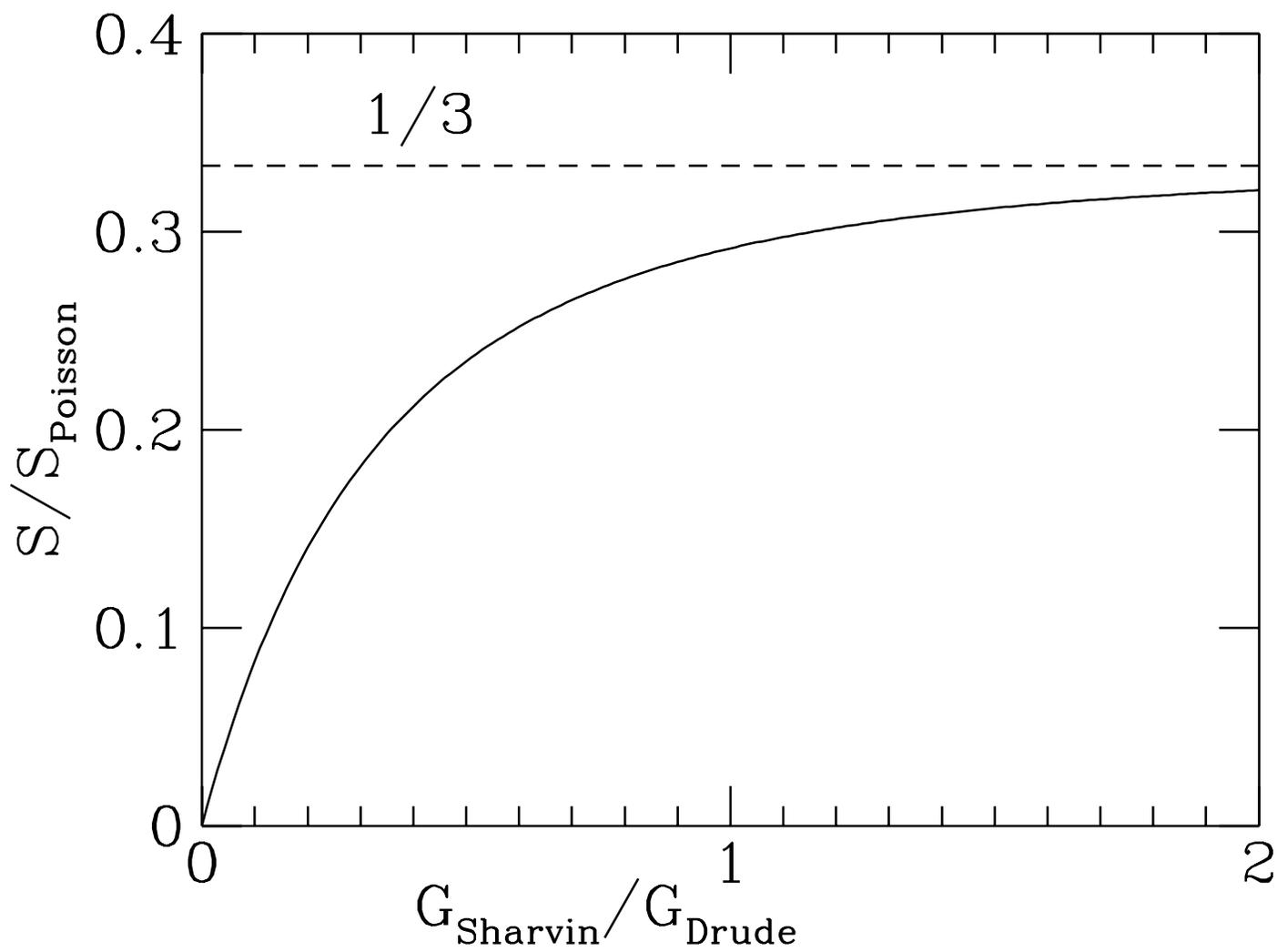
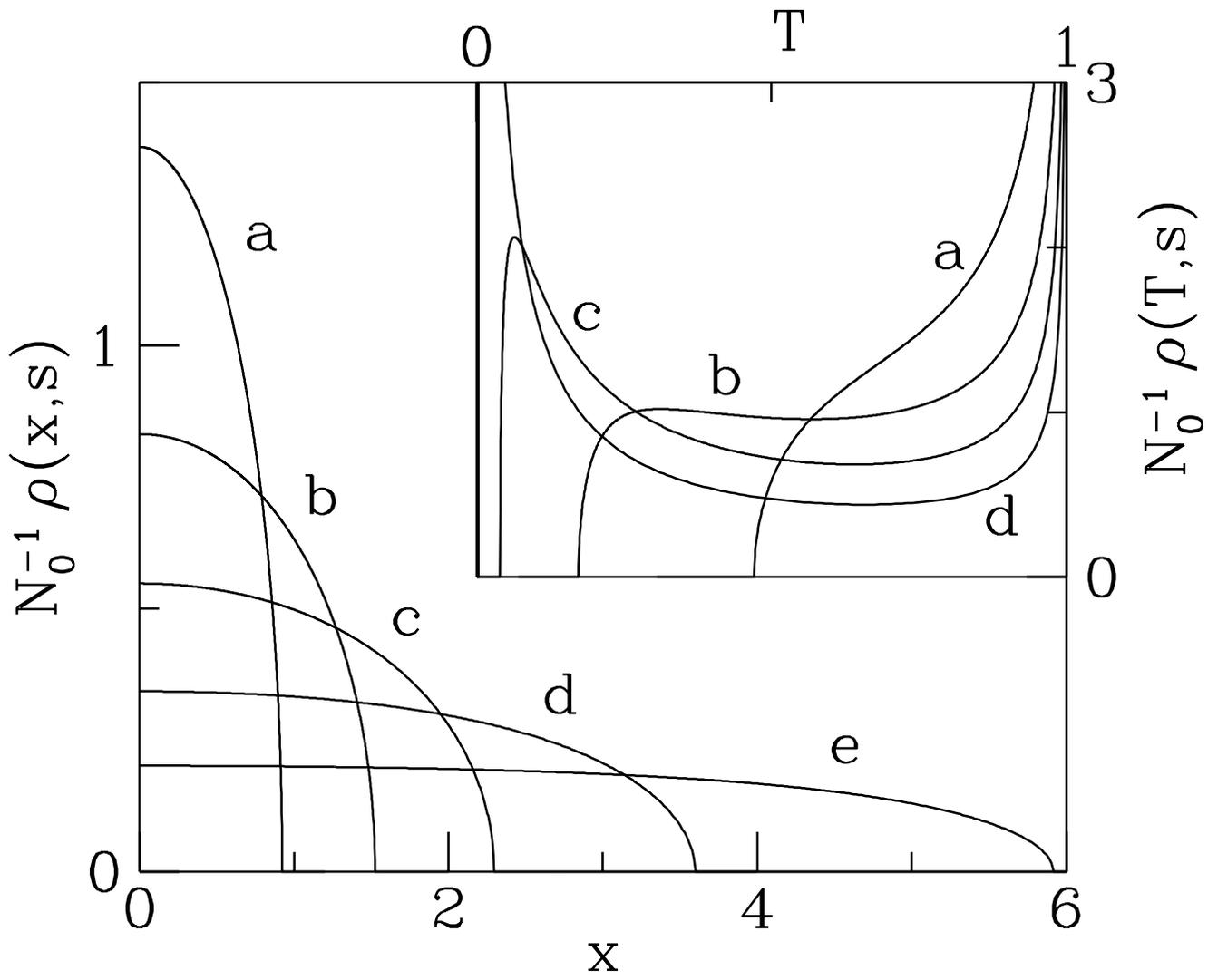
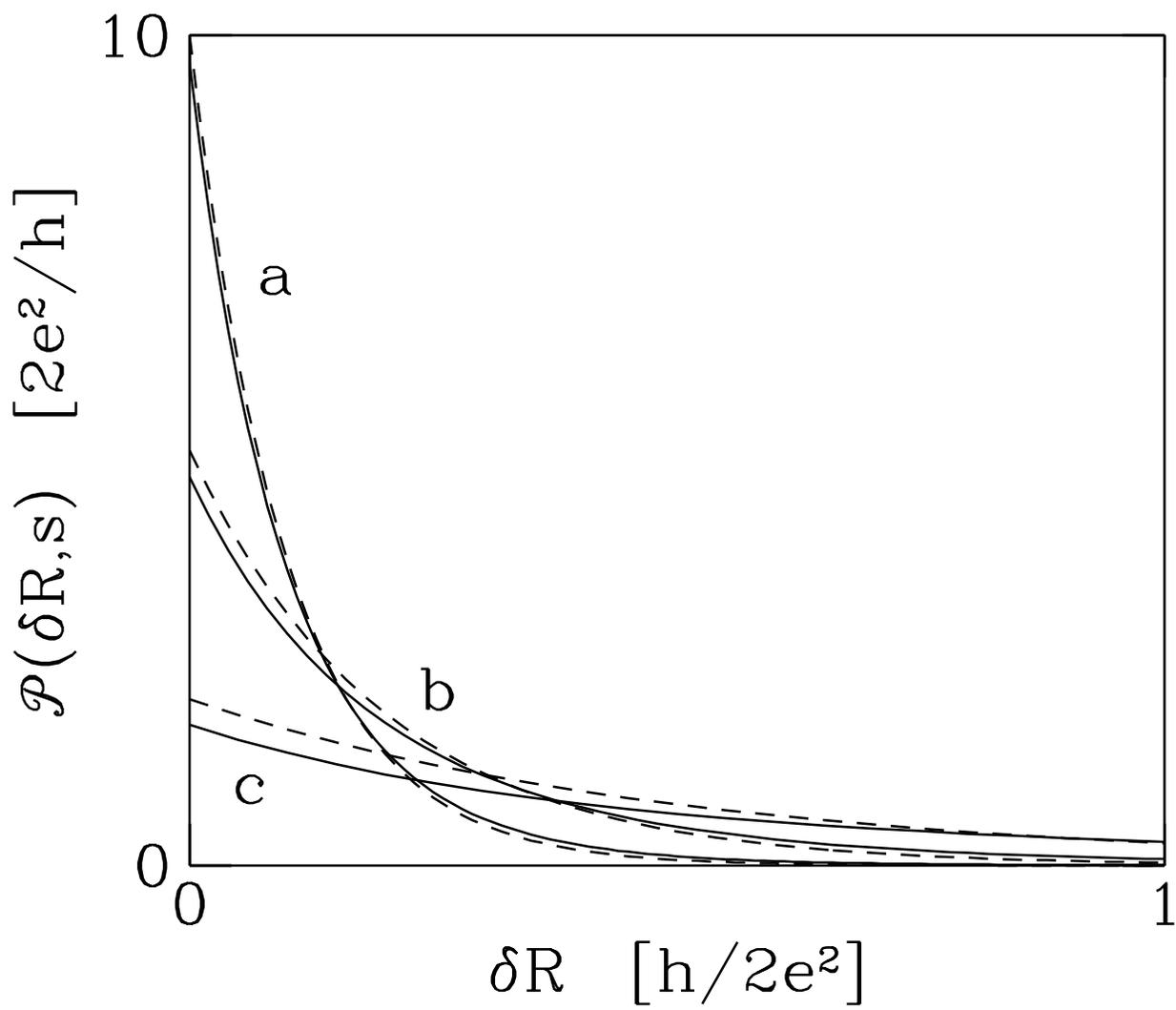
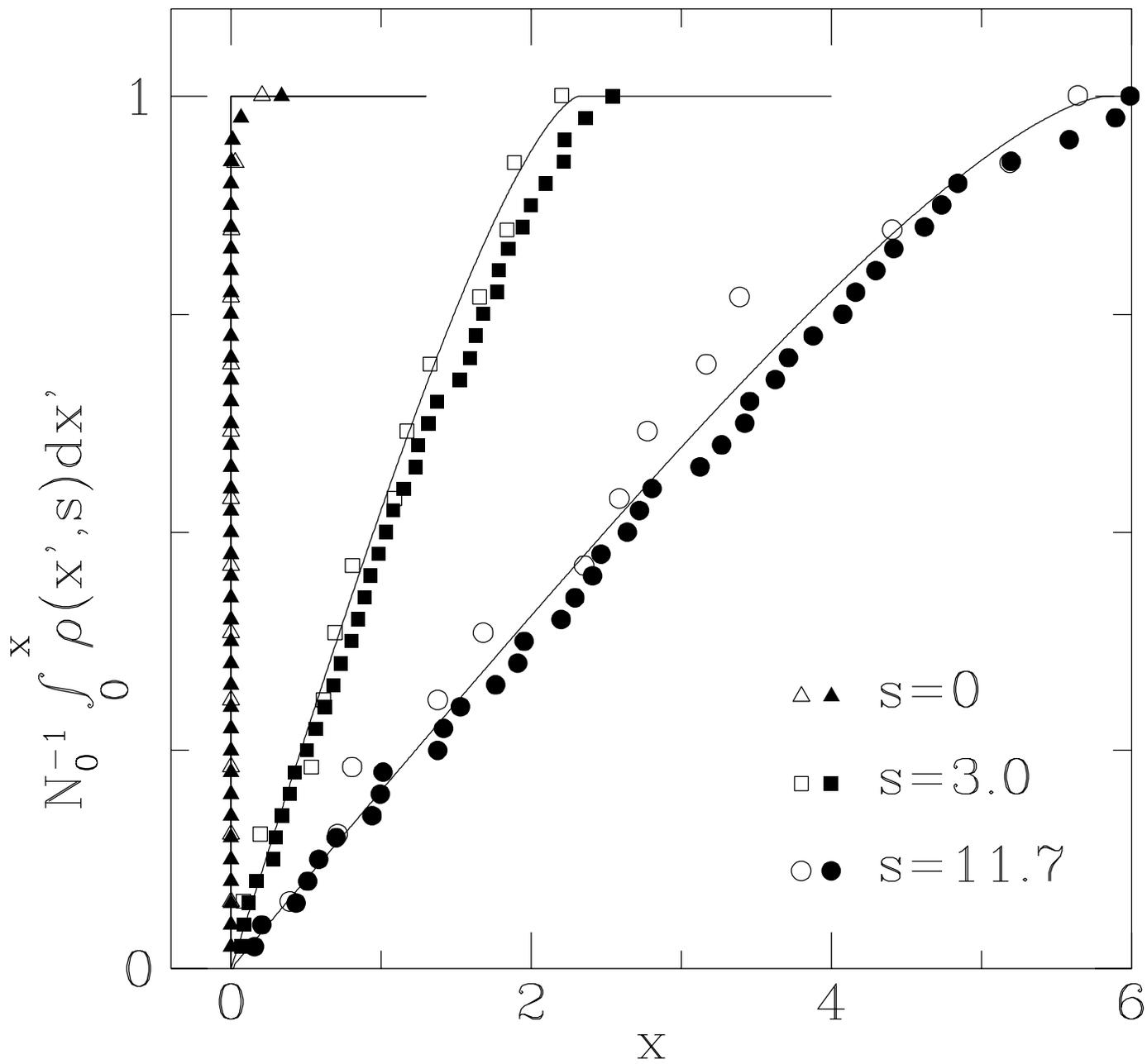
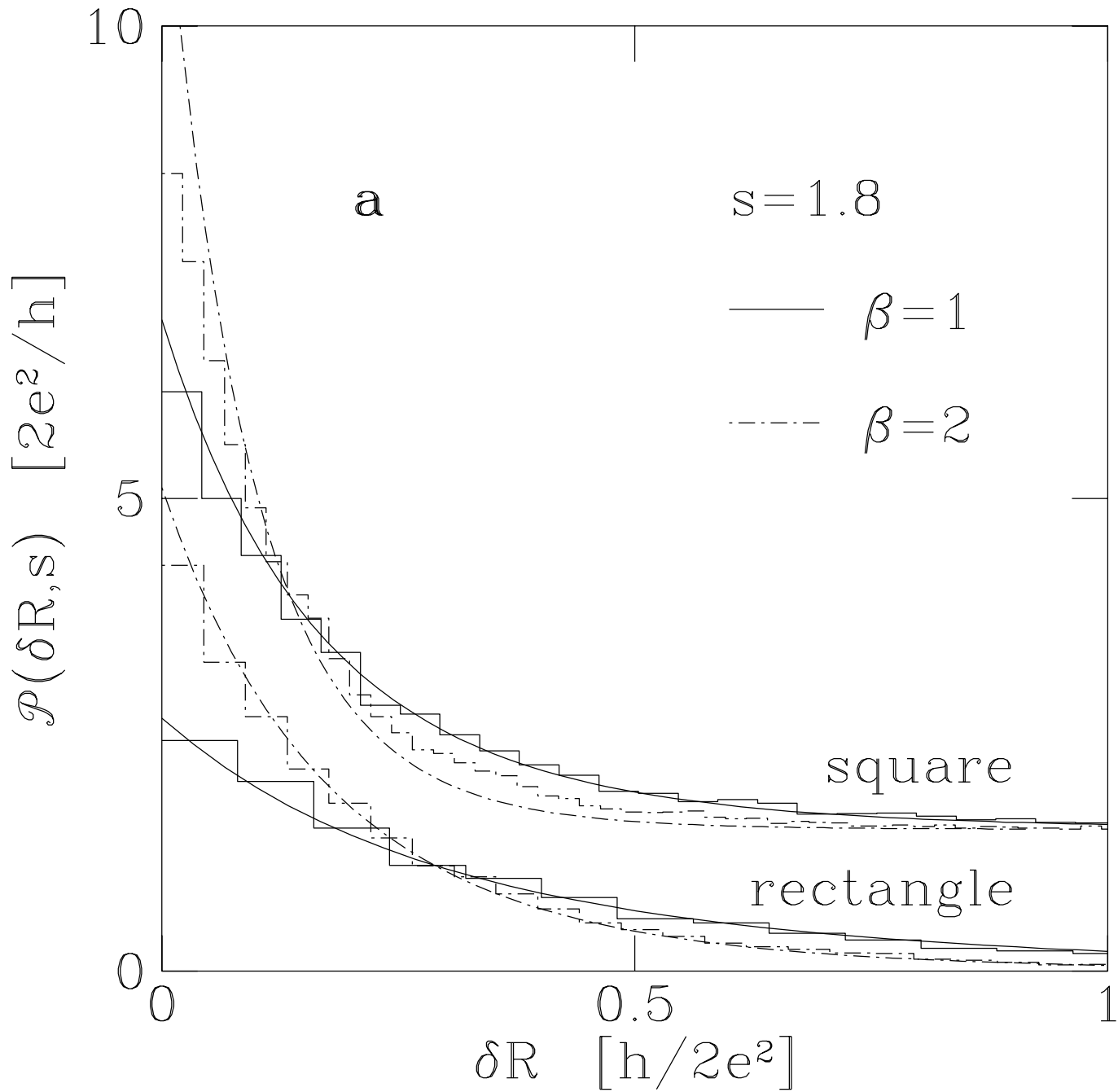
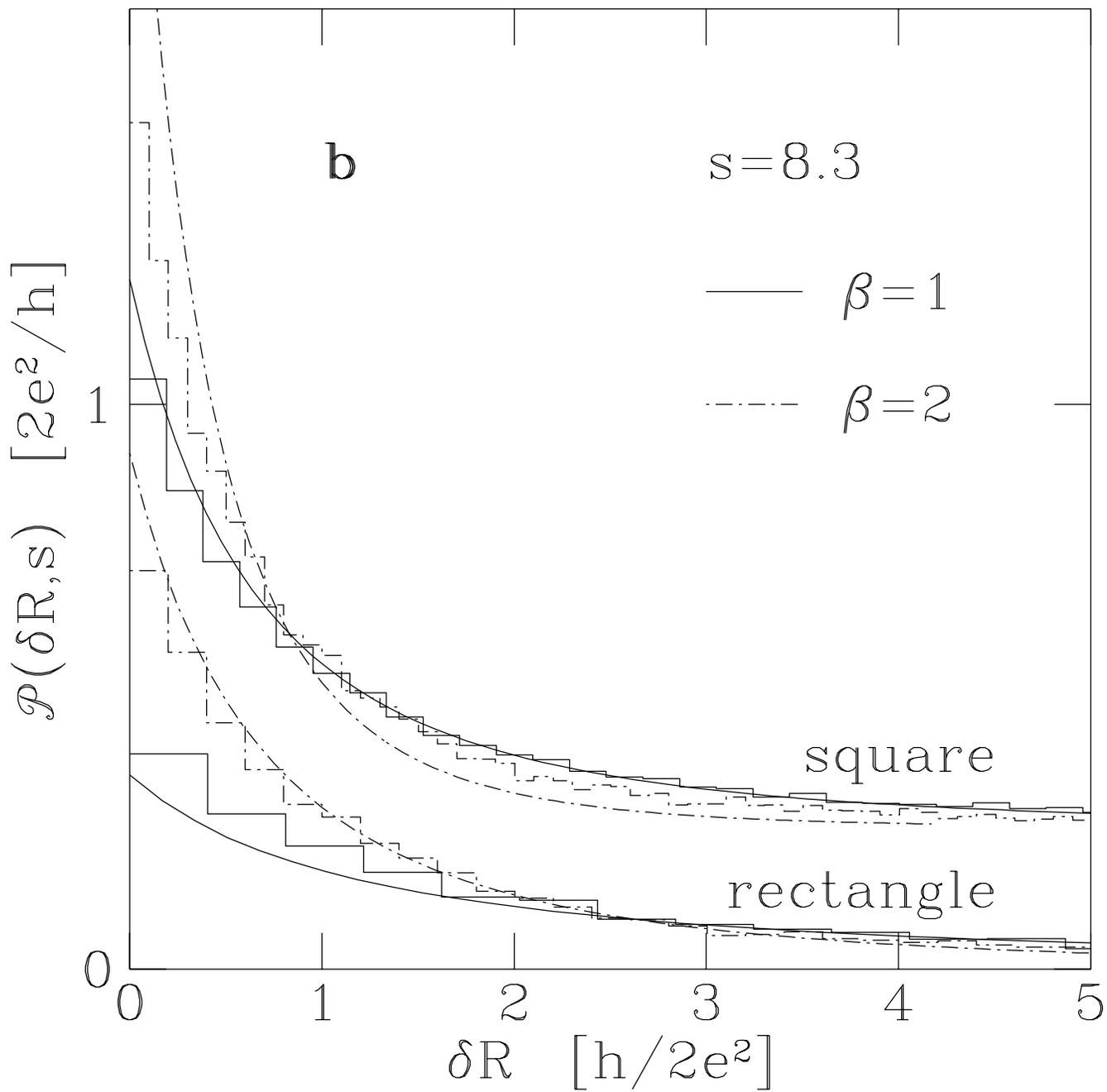
\end{document}